\documentclass[aps,prd,twocolumn,superscriptaddress,showpacs]{revtex4}
\usepackage[dvips]{graphicx}
\usepackage{amsmath,amssymb,times}
\usepackage{braket}

\newcommand{\bequ}{\begin{equation}}
\newcommand{\eequ}{\end{equation}}
\newcommand{\bea}{\begin{eqnarray}}
\newcommand{\eea}{\end{eqnarray}}

\DeclareSymbolFont{boldletters}{OML}{cmm} {b}{it}
\DeclareSymbolFontAlphabet{\mathbit}{boldletters}
\DeclareMathSymbol{\alpha}{\mathalpha}{letters}{"0B}
\DeclareMathSymbol{\beta}{\mathalpha}{letters}{"0C}
\DeclareMathSymbol{\gamma}{\mathalpha}{letters}{"0D}
\DeclareMathSymbol{\delta}{\mathalpha}{letters}{"0E}
\DeclareMathSymbol{\epsilon}{\mathalpha}{letters}{"0F}
\DeclareMathSymbol{\zeta}{\mathalpha}{letters}{"10}
\DeclareMathSymbol{\eta}{\mathalpha}{letters}{"11}
\DeclareMathSymbol{\theta}{\mathalpha}{letters}{"12}
\DeclareMathSymbol{\iota}{\mathalpha}{letters}{"13}
\DeclareMathSymbol{\kappa}{\mathalpha}{letters}{"14}
\DeclareMathSymbol{\lambda}{\mathalpha}{letters}{"15}
\DeclareMathSymbol{\mu}{\mathalpha}{letters}{"16}
\DeclareMathSymbol{\nu}{\mathalpha}{letters}{"17}
\DeclareMathSymbol{\xi}{\mathalpha}{letters}{"18}
\DeclareMathSymbol{\pi}{\mathalpha}{letters}{"19}
\DeclareMathSymbol{\rho}{\mathalpha}{letters}{"1A}
\DeclareMathSymbol{\sigma}{\mathalpha}{letters}{"1B}
\DeclareMathSymbol{\tau}{\mathalpha}{letters}{"1C}
\DeclareMathSymbol{\upsilon}{\mathalpha}{letters}{"1D}
\DeclareMathSymbol{\phi}{\mathalpha}{letters}{"1E}
\DeclareMathSymbol{\chi}{\mathalpha}{letters}{"1F}
\DeclareMathSymbol{\psi}{\mathalpha}{letters}{"20}
\DeclareMathSymbol{\omega}{\mathalpha}{letters}{"21}
\DeclareMathSymbol{\varepsilon}{\mathalpha}{letters}{"22}
\DeclareMathSymbol{\vartheta}{\mathalpha}{letters}{"23}
\DeclareMathSymbol{\varpi}{\mathalpha}{letters}{"24}
\DeclareMathSymbol{\varrho}{\mathalpha}{letters}{"25}
\DeclareMathSymbol{\varsigma}{\mathalpha}{letters}{"26}
\DeclareMathSymbol{\varphi}{\mathalpha}{letters}{"27}
\DeclareMathSymbol{\Gamma}{\mathalpha}{letters}{"00}
\DeclareMathSymbol{\Delta}{\mathalpha}{letters}{"01}
\DeclareMathSymbol{\Theta}{\mathalpha}{letters}{"02}
\DeclareMathSymbol{\Lambda}{\mathalpha}{letters}{"03}
\DeclareMathSymbol{\Xi}{\mathalpha}{letters}{"04}
\DeclareMathSymbol{\Pi}{\mathalpha}{letters}{"05}
\DeclareMathSymbol{\Sigma}{\mathalpha}{letters}{"06}
\DeclareMathSymbol{\Upsilon}{\mathalpha}{letters}{"07}
\DeclareMathSymbol{\Phi}{\mathalpha}{letters}{"08}
\DeclareMathSymbol{\Psi}{\mathalpha}{letters}{"09}
\DeclareMathSymbol{\Omega}{\mathalpha}{letters}{"0A}

\begin{document}
\title{
A practical solution to the sign problem at finite theta-vacuum angle
}

\author{Takahiro Sasaki}
\email[]{sasaki@phys.kyushu-u.ac.jp}
\affiliation{Department of Physics, Graduate School of Sciences, Kyushu University,
             Fukuoka 812-8581, Japan}

\author{Hiroaki Kouno}
\email[]{kounoh@cc.saga-u.ac.jp}
\affiliation{Department of Physics, Saga University,
             Saga 840-8502, Japan}

\author{Masanobu Yahiro}
\email[]{yahiro@phys.kyushu-u.ac.jp}
\affiliation{Department of Physics, Graduate School of Sciences, Kyushu University,
             Fukuoka 812-8581, Japan}

\date{\today}

\begin{abstract}
We propose a practical way of circumventing the sign problem 
in lattice QCD simulations with a theta-vacuum term. 
This method is the reweighting method for the QCD Lagrangian 
after the $U_{\rm A}(1)$ transformation. 
In the Lagrangian, the $P$-odd mass term as a cause of the sign problem 
is minimized. 
In order to find out a good reference system in the reweighting method, 
we estimate the average reweighting factor 
by using the two-flavor NJL model and eventually find a good reference system. 
\end{abstract}

\maketitle

\section{Introduction}

Phenomena based on strong interaction have shown that 
charge conjugation $C$, parity $P$ and time reversal ${\cal T}$ 
are good symmetries of nature. 
This means that quantum chromodynamics (QCD) should respect any combinations of the discrete symmetries. 
Among the discrete symmetries, the $CP$ symmetry is not necessarily respected in QCD due to the instanton solution~\cite{Instanton-75,tHooft}. 
The instanton solution allows the QCD Lagrangian ${\cal L}_{QCD}$ to have a $\theta$-vacuum term. The resulting Lagrangian is described as 
\bea
{\cal L}_{QCD}
&=&
\bar{q}_f(\gamma_{\nu}D_{\nu}+m_f)q_f
+\frac{1}{4}F^{a}_{\mu\nu}F^{a}_{\mu\nu} 
\nonumber \\
&~&-i\theta\frac{g^2}{64\pi^2}\epsilon_{\mu\nu\sigma\rho}F^{a}_{\mu\nu}F^{a}_{\sigma\rho} 
\label{QCD-1}
\eea
in Euclidean spacetime, 
where $F^{a}_{\mu\nu}$ is the field strength of gluon. 
The vacuum angle $\theta$ is a periodic variable with period $2\pi$. 
It was known to be an observable parameter~\cite{Theta76}. 
The Lagrangian ${\cal L}_{QCD}$ is invariant under the combination of the $P$ transformation and the parameter transformation $\theta \to -\theta$, 
indicating that the $P$ and $CP$ symmetries are preserved only at $\theta=0$ and $\pm \pi$; 
note that $\theta=-\pi$ is identical with $\theta=\pi$. 
In the $\theta$ vacuum, therefore, we must consider the $P$ and $CP$ violating interaction parameterized by $\theta$. 
Theoretically we can take any arbitrary value between $-\pi$ and $\pi$ for $\theta$. Nevertheless, 
it has been found from the measured neutron electric dipole moment~\cite{Baker} that $|\theta| < 10^{-9}$~\cite{Baluni,Crewther,Ohta}. 
Why is $\theta$ so small in zero temperature ($T$)? 
This long-standing puzzle is called the strong $CP$ problem; 
see for example Ref.~\cite{Vicari} for the review.

Around the deconfinement transition at $T=T_d$, 
there is a possibility that $P$-odd bubbles (metastable states) arise and thereby regions of nonzero $\theta$ are generated~\cite{Kharzeev:1998kz}. 
Thus $\theta$ can become a function depending on spacetime coordinates $(t,x)$. If $P$-odd bubbles are really produced at $T \approx \Lambda_{\mathrm QCD}$, 
$P$ and $CP$ symmetries can be violated locally in high-energy heavy-ion collisions or the early universe. 
This finite value of $\theta$ could be a new source of large $CP$ violation in the early universe and a crucial missing element for solving the puzzle of baryogenesis.

In the early stage of heavy-ion collision, the magnetic field is formed, 
and simultaneously the total number of particles plus antiparticles 
with right-handed helicity is deviated from that with left-handed helicity by the effective $\theta (t,x)$. 
In this situation, particles with right-handed helicity move opposite to antiparticles with right-handed helicity, and consequently an electromagnetic current is generated along the magnetic field. 
This is the so-called chiral magnetic effect~\cite{Kharzeev,KZ,FKW,Fukushima3}. The chiral magnetic effect may explain the charge separations observed in the recent STAR results~\cite{Abelev}. 
Hot QCD with nonzero $\theta$ is thus quite interesting.

For zero $T$ and zero quark-number chemical potential ($\mu$), some important properties are showed on $P$ symmetry. 
Vafa and Witten proved for $\theta=0$ that the vacuum is unique and conserves $P$ symmetry ~\cite{VW}. 
This theorem does not preclude the existence of $P$-odd bubbles.
At $\theta =\pi$, $P$ symmetry is also preserved as mentioned above, but it is spontaneously broken~\cite{Dashen,Witten}. 
The spontaneous violation of $P$ symmetry is called the Dashen mechanism~\cite{Dashen}. 
Although the mechanism is a nonperturbative phenomenon, the first-principle lattice QCD (LQCD) is not applicable for finite $\theta$ due to the sign problem. 
The mechanism at finite $T$ and/or finite $\mu$ was then investigated with effective models 
such as the chiral perturbation theory~\cite{VV,Smilga,Tytgat,ALS,Creutz,MZ}, 
the Nambu-Jona-Lasinio (NJL) model~\cite{FIK,Boer,Boomsma,Chatteriee}  
and the Polyakov-loop extended Nambu-Jona-Lasinio (PNJL) model~\cite{Kouno_CP,Sakai_CP,Sasaki}.

In the previous work~\cite{Sasaki}, we proposed a way of minimizing the sign problem on LQCD with finite $\theta$. 
The proposal is as follows. For simplicity, we consider two-flavor QCD. 
The QCD Lagrangian \eqref{QCD-1} is transformed into 
\begin{equation}
{\cal L}_{QCD}
=
\bar{q}'{\cal M}(\theta )q'
+\frac{1}{4}F^{a}_{\mu\nu}F^{a}_{\mu\nu}
\label{QCD-2}
\end{equation}
with 
\begin{eqnarray}
{\cal M}(\theta )&\equiv&\gamma_{\nu}D_{\nu}+m\cos{(\theta /2)}+mi\gamma_5\sin{(\theta /2)}
\end{eqnarray}
by using the $U_{\rm A}(1)$ transformation 
\begin{equation}
q=e^{i\gamma_5{\theta\over{4}}}q^\prime,
\label{UAtrans}
\end{equation}
where the quark field $q=(q_u,q_d)$ has been redefined by the new one 
$q^\prime$. The determinant ${\cal M}(\theta)$ satisfies 
\begin{eqnarray}
\det {\cal M}(\theta)
=
\left[
\det {\cal M}(-\theta)
\right]^{\ast},
\label{Sign-problem-1}
\end{eqnarray}
indicating that the sign problem is 
induced by the $P$-odd ($\theta$-odd) term, 
$mi\gamma_5 \sin{(\theta /2)}$. 
The difficulty of the sign problem is minimized in 
\eqref{QCD-2}, since the $P$-odd term with the light quark mass $m$ 
is much smaller than the dynamical quark mass of order $\Lambda_{\rm QCD}$. 
Actually, it was found that the $P$-even condensates 
$\sigma_f^\prime = \langle \bar{q}_f^\prime q_f^\prime \rangle$ 
is much larger than the $P$-odd condensates 
$\eta_f^\prime = \langle \bar{q}_f^\prime i\gamma_5q_f^\prime \rangle$. 
The $P$-even condensates little change even if 
the $\theta$-odd mass term is neglected. 
We then proposed the following reweighting method. 
The vacuum expectation value of operator ${\cal O}$ is calculated by 
\begin{eqnarray}
\braket{\cal O}
&=&
\int\mathcal{D}A
{\cal O}{\rm det}{\cal M}(\theta )e^{-S_g}
\\
&=&
\int\mathcal{D}A
{\cal O}'{\rm det}{\cal M}_{\rm ref}(\theta )e^{-S_g}
\label{reweighting-method-1}
\end{eqnarray}
with the gluon part $S_g$ of the QCD action and  
\begin{eqnarray}
{\cal O}'
&\equiv&
R(\theta ){\cal O},
\\
R(\theta )
&\equiv&
\frac{{\rm det}{\cal M}(\theta )}{{\rm det}{\cal M}_{\rm ref}(\theta )},
\end{eqnarray}
where $R(\theta )$ is the reweighting factor and $\det\mathcal{M}_{\rm ref}(\theta )$ is the Fermion determinant of the reference theory that has no sign problem. 
The simplest candidate of the reference theory is the theory 
in which the $\theta$-odd mass is neglected. 
We refer this reference theory to as reference A in this paper. 
As discussed in Ref.~\cite{Sasaki}, reference A may be 
a good reference theory for small and intermediate $\theta$, but not 
for large $\theta$ near $\pi$. In reference A, 
the limit of $\theta=\pi$ corresponds to the chiral limit 
that is hard for LQCD simulations to reach.

The expectation value of $R(\theta )$ in the reference theory is obtained by
\begin{equation}
\braket{R(\theta )}
=
\frac{Z}{Z_{\rm ref}}
\label{ab-R}
\end{equation}
where $Z$ ($Z_{\rm ref}$) is the partition function of the original (reference) theory.
The average reweighting factor $\braket{R(\theta )}$ is a good index 
for the reference theory to be good; 
the reference theory is good when $\braket{R(\theta )}=1$.

In this paper, we estimate $\braket{R(\theta)}$ with the 
NJL model in order to find out a good reference theory. 
We find that reference A is good only for small, so 
propose a good reference theory that satisfies $\braket{R(\theta)} \approx 1$.

This paper is organized as follows. 
In Sec.~\ref{Modelsetting}, we recapitulate the two-flavor NJL model 
and show how to calculate the pion mass and $\braket{R(\theta)}$
for the case of finite $\theta$. Numerical results are shown 
in Sec.~\ref{Numericalresults}. 
Section \ref{Summary} is devoted to summary.

\section{Model setting}
\label{Modelsetting}

The two-flavor NJL Lagrangian 
with the $\theta$-dependent anomaly term is obtained in Euclidean spacetime by
\begin{eqnarray}
\mathcal{L}
&=&
\bar{q}(\gamma_{\nu}\partial_{\nu}+m_0)q
-G_1\sum^{3}_{a=0}\left[
(\bar{q}\tau_aq)^2+(\bar{q}i\gamma_5\tau_aq)^2
\right]
\nonumber\\
&&~~~~
-8G_2\left[
e^{i\theta}{\rm det}\bar{q}_{\rm R}q_{\rm L}+e^{-i\theta}{\rm det}\bar{q}_{\rm L}q_{\rm R}
\right]
\end{eqnarray}
where $m_0$ is the current quark mass satisfying $m_0=m_u=m_d$ 
and $\tau_0$ and $\tau_a (a=1,2,3)$ are the $2\times 2$ unit and 
Pauli matrices in the flavor space, respectively. 
The parameter $G_1$ denotes the coupling constant 
of the scalar and pseudoscalar-type four-quark interaction, while 
$G_2$ stands for that of the Kobayashi-Maskawa-'t Hooft determinant interaction \cite{KMK,tHooft} where the matrix indices run in the flavor space.

Under the $U_A(1)$ transformation \eqref{UAtrans}, 
the quark-antiquark condensates, 
$\{\phi_i\} \equiv \{\sigma, \eta, a_i, \pi_i\}$, 
are transformed as
\begin{eqnarray}
&&
\sigma
\equiv
\bar{q}q
=
{\rm cos}(\theta /2)\sigma'+{\rm sin}(\theta /2)\eta',
\\
&&
\eta
\equiv
\bar{q}i\gamma_5q
=
-{\rm sin}(\theta /2)\sigma'+{\rm cos}(\theta /2)\eta',
\\
&&
a_i
\equiv
\bar{q}\tau_iq
=
{\rm cos}(\theta /2)a_i'+{\rm sin}(\theta /2)\pi_i',
\\
&&
\pi_i
\equiv
\bar{q}i\gamma_5\tau_iq
=
-{\rm sin}(\theta /2)a_i'+{\rm cos}(\theta /2)\pi_i',
\end{eqnarray}
where the condensates $\{\phi_i'\} \equiv \{\sigma', \eta', a_i', \pi_i'\}$ 
are defined by the same form as $\{\phi_i\}$ but $q$ is replaced by $q'$. 
The Lagrangian density is then rewritten with $q'$ as
\begin{eqnarray}
\mathcal{L}
&=&
\bar{q}'(\gamma_{\nu}\partial_{\nu}+m(\theta ))q'
-G_1\sum^{3}_{a=0}\left[
(\bar{q}'\tau_aq')^2+(\bar{q}'i\gamma_5\tau_aq')^2
\right]
\nonumber\\
&&~~~~
-8G_2\left[
{\rm det}\bar{q}'_{\rm R}q'_{\rm L}+{\rm det}\bar{q}'_{\rm L}q'_{\rm R}
\right]
\\
&=&
\bar{q}'(\gamma_{\nu}\partial_{\nu}+m(\theta ))q'
-G_+\left[
(\bar{q}'q')^2+(\bar{q}'i\gamma_5\vec{\tau}q')^2
\right]
\nonumber\\
&&~~~~
-G_-\left[
(\bar{q}'\vec{\tau}q')^2+(\bar{q}'i\gamma_5q')^2 
\right] ,
\label{NJL-3}
\end{eqnarray}
where $G_{\pm}=G_1\pm G_2$ and 
\begin{eqnarray}
m(\theta )
&=&
m_0{\rm cos}(\theta /2)+m_0i\gamma_5{\rm sin}(\theta /2).
\end{eqnarray}
Making the mean field approximation and the path integral 
over the quark field, 
one can obtain the thermodynamic potential $\Omega$ (per volume) 
for finite $T$:
\begin{equation}
\Omega
=
U-4N_{\rm c}\sum_{\pm}
\int\frac{d^3p}{(2\pi )^3}
\Bigl[
E_{\pm}
+\frac{1}{\beta}{\rm ln}\left[ 1+e^{-\beta E_{\pm}}\right]
\Bigr] 
\end{equation}
with 
\begin{eqnarray}
&&
E_{\pm}=\sqrt{\vec{p}^{~2}+C\pm 2\sqrt{D}}
,\\
&&
C=M^2+N^2+A^2+P^2
,\\
&&
D=(M\vec{A}+N\vec{P})^2+(\vec{A}\times \vec{P})^2
\ge 0
,\\
&&
M=
m_0{\rm cos}(\theta /2)-2G_+\sigma'
,\\
&&
N=
m_0{\rm sin}(\theta /2)-2G_-\eta'
,\\
&&
\vec{A}=-2G_-\vec{a}',~~\vec{P}=-2G_+\vec{\pi}'
,\\
&&
A=\sqrt{\vec{A}\cdot\vec{A}},~~P=\sqrt{\vec{P}\cdot\vec{P}}
,\\
&&
U=
G_+(\sigma^{\prime 2}+\vec{\pi}^{\prime 2})
+
G_-(\eta^{\prime 2}+\vec{a}^{\prime 2}),
\end{eqnarray}
where the momentum integral is regularized by 
the three-dimensional momentum cutoff $\Lambda$. 
Following Refs. \cite{Boer,Boomsma}, we introduce a parameter $c$ 
as $G_1=(1-c)G_+$ and $G_2=cG_+$, where $0\le c\le 0.5$ and $G_+>0$. 
The present model thus has four parameters 
of $m_0$, $\lambda$, $G_+$ and $c$. 
Assuming $m_0=5.5$ MeV, we have determined $\Lambda$ and $G_+$ 
from the pion decay constant $f_{\pi}=93$ MeV 
and the pion mass $M_{\pi}=138$ MeV at vacuum. 
Although $c$ is an unknown parameter, we set $c=0.2$ here, 
since it is known from the model analysis on the $\eta -\eta'$ splitting
that $c \approx 0.2$ is favorable \cite{Frank}.

For finite $\theta$, parity is broken explicitly, so 
it is not a good quantum number anymore. 
Hence $P$-even and $P$-odd modes are mixed with each other for each meson. 
The ``pion" mass $\tilde{M}_{\pi}$ is defined 
by the lowest pole mass of the inverse propagator in 
the isovector channel. 
It agrees with the ordinary pion mass when $\theta =0$.
Under the random phase approximation~\cite{Hansen}, 
the inverse propagator is described by 
\begin{equation}
{\rm det}[1-2G\Pi (\tilde{M}_{\pi}^2)]=0 ,
\end{equation}
where
\begin{eqnarray}
G&=&
\left(
\begin{array}{cc}
G_-&0\\
0&G_+
\end{array}
\right) ,
\\
\Pi (q^2)
&=&
\left(
\begin{array}{cc}
\Pi^{SS}(q^2)&\Pi^{SP}(q^2)\\
\Pi^{PS}(q^2)&\Pi^{PP}(q^2)
\end{array}
\right)
\end{eqnarray}
with
\begin{eqnarray}
\Pi^{PP}
&=&
4N_fN_cI_1 - 2N_fN_c(q^2-4N^2)I_2(q^2),
\\
\Pi^{SS}
&=&
4N_fN_cI_1 - 2N_fN_c(q^2-4M^2)I_2(q^2),
\\
\Pi^{SP}
&=&
\Pi^{PS}
=
-8N_fN_cMNI_2(q^2),
\\
I_1
&=&
\int\frac{d^4p}{(2\pi )^4}\frac{1}{p^2-M^2-N^2},
\\
I_2(q^2)
&=&
\int\frac{d^4p}{(2\pi )^4}
\prod_{\pm}
\frac{1}{
\left[
(p\pm q/2)^2-M^2-N^2
\right]
}.
\end{eqnarray}
In this form, we can set $\vec{a}'=\vec{\pi}'=0$, since 
we do not consider the isospin chemical potential.

Applying the saddle-point approximation to the path integral 
in the partition function, one can get 
\begin{equation}
\braket{R(\theta )}
\approx
\sqrt{\frac{{\rm det}H_{\rm ref}}{{\rm det}H}}e^{-\beta V(\Omega -\Omega_{\rm ref})}
\end{equation}
where $\beta =1/T$, 
$\Omega$ ($\Omega_{\rm ref}$) is the thermodynamic potential 
at the mean-field level in the original (reference) theory 
and $H$ ($H_{\rm ref}$) is the Hessian matrix in the original (reference) theory 
defined by \cite{Andersen,Sakai_sign}
\begin{eqnarray}
H_{ij}
&=&
\frac{\partial^2\Omega}{\partial\phi_i'\partial\phi_j'}.
\end{eqnarray}
For later convenience, the average reweighting factor $\braket{R(\theta )}$ 
is divided into two factors $R_A$ and $R_B$:
\begin{eqnarray}
\braket{R(\theta )}&=&R_AR_B
\end{eqnarray}
with 
\begin{eqnarray}
R_A&=&\sqrt{\frac{{\rm det}H_{\rm ref}}{{\rm det}H}},\\
R_B&=&e^{-\beta V(\Omega -\Omega_{\rm ref})}.
\end{eqnarray}
For the $N_x^3 \times N_{\tau}$ lattice, 
the four-dimensional volume $\beta V$ is obtained by 
\begin{equation}
\beta V=\left( \frac{N_x}{N_{\tau}} \right)^3 \frac{1}{T^4}.
\end{equation}
Here we consider $N_{x}/N_{\tau}=4$ as a typical example, following 
Refs.~\cite{Andersen,Sakai_sign}. 

We consider the following reference theory that has no sign problem:
\begin{eqnarray}
\mathcal{L}
&=&
\bar{q}'(\gamma_{\nu}\partial_{\nu}+m_{\rm ref}(\theta ))q'
-G_+\left[
(\bar{q}'q')^2+(\bar{q}'i\gamma_5\vec{\tau}q')^2
\right]
\nonumber\\
&&~~~~
-G_-\left[
(\bar{q}'\vec{\tau}q')^2+(\bar{q}'i\gamma_5q')^2
\right].
\end{eqnarray}
Here $m_{\rm ref}(\theta )$ is $\theta$-even mass defined below. 
We consider three examples as $m_{\rm ref}(\theta )$. 

\begin{description}
\item[A.] 
The first example is reference A defined by 
\begin{equation}
m_{\rm ref}(\theta ) \equiv m_{\rm A}(\theta) =
m_0{\rm cos}(\theta /2).
\end{equation}
In this case, the $P$-odd mass is simply neglected from the original 
Lagrangian \eqref{NJL-3}. 

\item[B.] 
The second example is reference B defined by 
\begin{eqnarray}
m_{\rm ref}(\theta ) &{\equiv}& 
m_{\rm B}(\theta ) \nonumber \\ 
&=&
m_0{\rm cos}(\theta /2)
+
\frac{1}{\alpha}\left\{m_0{\rm sin}(\theta /2)\right\}^2.
\end{eqnarray}
In this case, we have added the $m_0^2$-order correction 
due to the $P$-odd quark mass. 
Here $\alpha$ is a parameter with mass dimension, so we simply 
choose $\alpha =M_{\pi}$. The coefficient of the correction term 
is $m_0^2/M_{\pi}=0.129$~MeV. 

\item[C.] 
The third case is reference C defined by 
\begin{eqnarray}
m_{\rm ref}(\theta ) &{\equiv}& 
m_{\rm C}(\theta ) \nonumber \\ 
&=&
m_0{\rm cos}(\theta /2)
+
\frac{m_0M_{\pi}^2}{M_{\eta'}^2}
{\rm sin}^2(\theta /2),
\end{eqnarray}
This case also has the $m_0^2$-order correction, but $\alpha$ is different 
from reference B. The coefficient of the correction term is 
$m_0 M_{\pi}^2/M_{\eta'}^2=0.114$~MeV. 
\end{description}

Reference C is justified as follows. 
The pion mass $\tilde{M}_{\pi}(\theta)$ at finite $\theta$  
is estimated from the chiral Lagrangian as~\cite{MZ}:
\begin{eqnarray}
\tilde{M}_{\pi}^2(\theta)
&=&
\frac{m_0|\sigma_0|}{f_{\pi}^2}|{\rm cos}(\theta /2)|
+
\frac{2l_7m_0^2\sigma_0^2}{f^6_{\pi}}
{\rm sin}^2(\theta /2),
\label{mpi-largeN}
\end{eqnarray}
where $\sigma_0$ is the chiral condensate at $T=\theta =0$ and 
the coefficient $l_7$ is evaluated by the $1/N_c$ expansion as 
\bea
l_7
&\approx&
\frac{f_{\pi}^2}{2M_{\eta'}^2}.
\eea
The right-hand side of \eqref{mpi-largeN} is reduced to
\begin{equation}
\tilde{M}_{\pi}^2(\theta)
=
\frac{|\sigma_0|}{f_{\pi}^2}
\left[
m_0|{\rm cos}(\theta /2)|
+
\frac{m_0M_{\pi}^2}{M_{\eta'}^2}
{\rm sin}^2(\theta /2)
\right].
\label{mpi-largeN-2}
\end{equation}
Equation \eqref{mpi-largeN-2} supports \eqref{mpi-largeN}.

\section{Numerical results}
\label{Numericalresults}
%
\begin{figure}[t]
\begin{center}
\hspace{-10pt}
 \includegraphics[height=0.18\textheight,bb=90 50 232 201,clip]{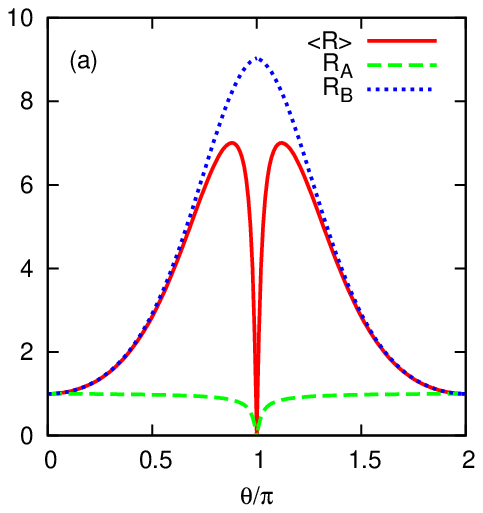}
 \includegraphics[height=0.18\textheight,bb=75 50 232 201,clip]{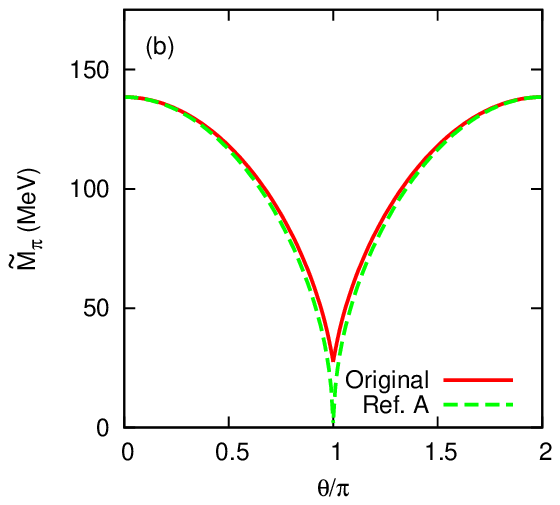}
\vspace{-10pt}
\end{center}
\caption{
$\theta$ dependence of (a) the average reweighting factor 
and (b) $\tilde{M}_{\pi}$ at $T=100$ MeV for the case of reference A.
}
\label{case1}
\end{figure}
%
%
\begin{figure}[t]
\begin{center}
\hspace{-10pt}
 \includegraphics[height=0.18\textheight,bb=90 50 232 201,clip]{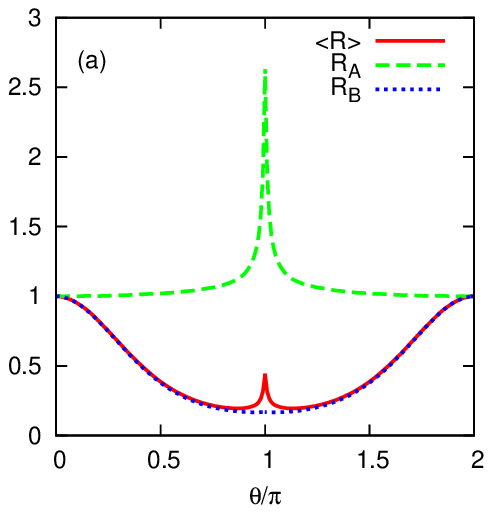}
 \includegraphics[height=0.18\textheight,bb=75 50 232 201,clip]{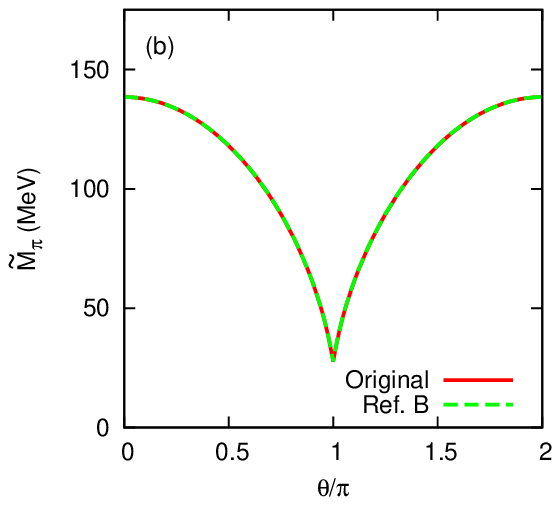}
\vspace{-10pt}
\end{center}
\caption{
$\theta$ dependence of (a) the average reweighting factor 
and (b) $\tilde{M}_{\pi}$ at $T=100$ MeV for the case of reference B.
}
\label{case2}
\end{figure}
%
%
\begin{figure}[t]
\begin{center}
\hspace{-10pt}
 \includegraphics[height=0.18\textheight,bb=90 50 232 201,clip]{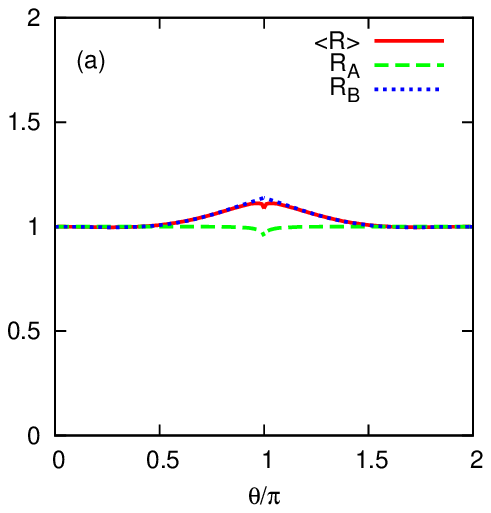}
 \includegraphics[height=0.18\textheight,bb=75 50 232 201,clip]{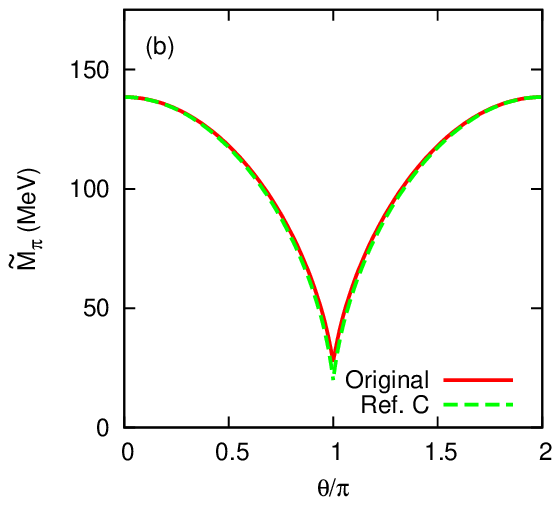}
\vspace{-10pt}
\end{center}
\caption{
$\theta$ dependence of (a) the average reweighting factor 
and (b) $\tilde{M}_{\pi}$ at $T=100$ MeV for the case of reference C.
}
\label{case3}
\end{figure}
%
If some reference system satisfies the condition that 
$\braket{R(\theta )} \approx 1$, one can say that the reference system is 
good. As a typical example of the condition, we consider the case of 
$0.5 \lesssim \braket{R(\theta )} \lesssim 2$. This condition seems to be 
the minimum requirement. 
The discussion made below is not changed qualitatively, even if one takes 
a stronger condition.

First we consider reference A. 
Figure \ref{case1}(a) shows $\theta$ dependence of $\braket{R(\theta )}$ 
at $T=100$ MeV.
The solid line stands for $\braket{R(\theta )}$, while 
the dashed (dotted) line corresponds to $R_{\rm A}$ ($R_{\rm B}$).
This temperature is lower than the chiral transition temperature 
in the original theory that is 
$206$ MeV at $\theta=0$ and $194$ MeV at $\theta=\pi$.
As $\theta$ increases from zero, $\braket{R(\theta )}$ also increases and 
exceeds 2 at $\theta \approx 1.2$. Reference A is thus good for 
$\theta \lesssim 1.2$. The increase of $\braket{R(\theta )}$ stems from that of $R_{\rm B}$ that depends on $T$. This means that 
the reliable $\theta$ region in which 
$0.5 \lesssim \braket{R(\theta )} \lesssim 2$ 
becomes large as $T$ increases.

Figure \ref{case1}(b) shows $\theta$ dependence of $\tilde{M}_{\pi}$ 
at $T=100$ MeV. The solid (dashed) line denotes $\tilde{M}_{\pi}$ 
in the original (reference A) system. 
At $\theta =\pi$, $\tilde{M}_{\pi}$ is finite 
in the original system, but zero in reference A. 
As a consequence of this property, $R_{\rm A}$ and $\braket{R(\theta )}$ 
vanish at $\theta =\pi$; see Fig. \ref{case1}(a). 
This indicates that reference A breaks down at $\theta =\pi$, 
independently of $T$.

The same analysis is made for reference B in Fig. \ref{case2}. 
As shown in panel (b), $\tilde{M}_{\pi}$ in reference B well reproduces 
that in the original theory for any $\theta$. 
As shown in panel (a), however, the reliable $\theta$ region 
in which $0.5 \lesssim \braket{R(\theta )} \lesssim 2$ is located only at $\theta \lesssim 1.3$. 
Therefore reference B is still not good.

Finally we consider reference C. 
As shown in Fig. \ref{case3}(b), $\tilde{M}_{\pi}$ in reference C 
well simulates that of the original theory at small and intermediate 
$\theta$, but the former slightly underestimates the latter 
at large $\theta$ around $\pi$. 
As shown in Fig. \ref{case3}(b), 
however, $\braket{R(\theta )}$ satisfies 
the condition $0.5 \lesssim \braket{R(\theta )} \lesssim 2$ for all $\theta$. 
Therefore we can think that reference C is a good reference system 
for any $\theta$.
This is true for any temperature larger than 100~MeV. 

\section{Summary}
\label{Summary}
We have proposed a practical way of circumventing the sign problem 
in LQCD simulations with finite $\theta$. 
This method is the reweighting method for 
the transformed Lagrangian \eqref{QCD-2}. In the Lagrangian, 
The sign problem is minimized, 
since the $P$-odd mass is much smaller than 
the dynamical quark mass of order $\Lambda_{\rm QCD}$. 
Another key is to find out which kind of reference system satisfies 
the condition $\braket{R(\theta )} \approx 1$. 
For this purpose, we have estimated $\braket{R(\theta )}$ 
by using the two-flavor NJL model and eventually found that 
reference C is a good reference system in the reweighting method.

\begin{acknowledgments}
The authors thank A. Nakamura and P. de Forcrand for useful discussions. 
H.K. also thanks M. Imachi, H. Yoneyama, H. Aoki, and M. Tachibana for useful discussions. 
T.S. is supported by JSPS.
\end{acknowledgments}



\begin{thebibliography}{19}
\expandafter\ifx\csname natexlab\endcsname\relax\def\natexlab#1{#1}\fi
\expandafter\ifx\csname bibnamefont\endcsname\relax
  \def\bibnamefont#1{#1}\fi
\expandafter\ifx\csname bibfnamefont\endcsname\relax
  \def\bibfnamefont#1{#1}\fi
\expandafter\ifx\csname citenamefont\endcsname\relax
  \def\citenamefont#1{#1}\fi
\expandafter\ifx\csname url\endcsname\relax
  \def\url#1{\texttt{#1}}\fi
\expandafter\ifx\csname urlprefix\endcsname\relax\def\urlprefix{URL }\fi
\providecommand{\bibinfo}[2]{#2}
\providecommand{\eprint}[2][]{\url{#2}}


%
\bibitem[{\citenamefont{Belavin}(1975)}]{Instanton-75}
\bibinfo{author}{\bibfnamefont{A.A.}~\bibnamefont{Belavin}},
\bibinfo{author}{\bibfnamefont{A.M.}~\bibnamefont{Polyakov}},
\bibinfo{author}{\bibfnamefont{A.S.}~\bibnamefont{Schwartz}},
\bibnamefont{and}
\bibinfo{author}{\bibfnamefont{Yu.S.}~\bibnamefont{Tyupkin}},
\bibinfo{journal}{Phys. Lett.\ B} \textbf{\bibinfo{volume}{59}},
\bibinfo{pages}{85} (\bibinfo{year}{1975}).
%
\bibitem[{\citenamefont{'t Hooft}(1976)}]{tHooft}
\bibinfo{author}{\bibfnamefont{G.}~\bibnamefont{'t Hooft}},
\bibinfo{journal}{Phys. Rev.\ Lett.} \textbf{\bibinfo{volume}{37}},
\bibinfo{pages}{8} (\bibinfo{year}{1976});
\bibinfo{journal}{Phys. Rev.\ D} \textbf{\bibinfo{volume}{14}},
\bibinfo{pages}{3432} (\bibinfo{year}{1976});
\textbf{\bibinfo{volume}{18}},
\bibinfo{pages}{2199(E)} (\bibinfo{year}{1978}).
%
\bibitem{Theta76}
\bibinfo{author}{\bibfnamefont{C.}~\bibfnamefont{G.}~\bibnamefont{Callan}}, 
\bibinfo{author}{\bibfnamefont{R.}~\bibfnamefont{F.}~\bibnamefont{Dashen}}, 
\bibnamefont{and}
\bibinfo{author}{\bibfnamefont{D.}~\bibfnamefont{J.}~\bibnamefont{Gross}}, 
\bibinfo{journal}{Phys. Lett.} \textbf{B63},
\bibinfo{pages}{334} (\bibinfo{year}{1976}); 
\bibinfo{author}{\bibfnamefont{J.}~\bibnamefont{Jackiw}},
\bibnamefont{and}
\bibinfo{author}{\bibfnamefont{C.}~\bibnamefont{Rabbi}},
\bibinfo{journal}{Phys. Rev. Lett.} \textbf{\bibinfo{volume}{37}},
\bibinfo{pages}{172} (\bibinfo{year}{1976}). 
%
\bibitem{Baker}
\bibinfo{author}{\bibfnamefont{C.}~\bibfnamefont{A.}~\bibnamefont{Baker}},
et al.,
\bibinfo{journal}{Phys. Rev. Lett.} \textbf{\bibinfo{volume}{97}},
\bibinfo{pages}{131801} (\bibinfo{year}{2006}). 
%
\bibitem{Baluni}
V. Baluni, Phys. Rev. {\bf D19}, 2227(1979).
%
\bibitem{Crewther}
R. Crewther, P. di Vecchia, G. Veneziano and E. Witten, Phys. Lett. 
{\bf 88B}, 123(1979);{91B}, 487(1980)(E). 
%
\bibitem{Ohta}
\bibinfo{author}{\bibfnamefont{K.}~\bibnamefont{Kawarabayashi}}
\bibnamefont{and}
\bibinfo{author}{\bibfnamefont{N.}~\bibnamefont{Ohta}},
\bibinfo{journal}{Nucl. \ Phys. } \textbf{\bibinfo{volume}{B175}},
\bibinfo{pages}{477} (\bibinfo{year}{1980}); 
\bibinfo{journal}{Prog. Theor. Phys.} \textbf{\bibinfo{volume}{66}},
\bibinfo{pages}{1789} (\bibinfo{year}{1981}); 
\bibinfo{author}{\bibfnamefont{N.}~\bibnamefont{Ohta}},
\bibinfo{journal}{Prog. Theor. Phys.} \textbf{\bibinfo{volume}{66}},
\bibinfo{pages}{1408} (\bibinfo{year}{1981}); 
[Erratum-ibid.\  {\bf 67} (1982) 993].
%
\bibitem{Vicari}
\bibinfo{author}{\bibfnamefont{E.}~\bibnamefont{Vicari}} 
\bibnamefont{and}
\bibinfo{author}{\bibfnamefont{H.}~\bibnamefont{Panagopoulos}},
\bibinfo{journal}{Phys. Rept. } \textbf{\bibinfo{volume}{470}},
\bibinfo{pages}{93} (\bibinfo{year}{2009}). 
%
\bibitem{Kharzeev:1998kz} 
D.~Kharzeev, R.~D.~Pisarski and M.~H.~G.~Tytgat,
Phys.\ Rev.\ Lett.\  {\bf 81}, 512 (1998),  
[hep-ph/9804221] and [hep-ph/0012012].  
%
\bibitem[{\citenamefont{Kharzeev and Zhitnitsky}(2007)}]{KZ}
\bibinfo{author}{\bibfnamefont{D.}~\bibnamefont{Kharzeev}}, 
\bibnamefont{and}
\bibinfo{author}{\bibfnamefont{A.}~\bibnamefont{Zhitnitsky}},
\bibinfo{journal}{Nucl. \ Phys. A} \textbf{\bibinfo{volume}{797}},
\bibinfo{pages}{67} (\bibinfo{year}{2007}). 
%
\bibitem{Kharzeev}
\bibinfo{author}{\bibfnamefont{D.}~\bibnamefont{Kharzeev}}, 
\bibinfo{journal}{Phys. Lett.\ B} \textbf{\bibinfo{volume}{633}},
\bibinfo{pages}{260} (\bibinfo{year}{2006}); 
\bibinfo{author}{\bibfnamefont{D.}~\bibnamefont{Kharzeev}}, 
\bibinfo{author}{\bibfnamefont{L.}~\bibfnamefont{D.}~\bibnamefont{McLerran}}, 
\bibnamefont{and}
\bibinfo{author}{\bibfnamefont{H.}~\bibfnamefont{J.}~\bibnamefont{Warringa}},
\bibinfo{journal}{Nucl. Phys. A} \textbf{\bibinfo{volume}{803}},
\bibinfo{pages}{227} (\bibinfo{year}{2008}). 
%
\bibitem{FKW}
\bibinfo{author}{\bibfnamefont{K.}~\bibnamefont{Fukushima}}, 
\bibinfo{author}{\bibfnamefont{D.}~\bibfnamefont{E.}~\bibnamefont{Kharzeev}}, 
\bibnamefont{and}
\bibinfo{author}{\bibfnamefont{H.}~\bibfnamefont{J.}~\bibnamefont{Warringa}},
\bibinfo{journal}{Phys. Rev. D} \textbf{\bibinfo{volume}{78}}, 
\bibinfo{pages}{074033} (\bibinfo{year}{2008}). 
%
\bibitem[{\citenamefont{Fukushima et al.}(2010)}]{Fukushima3}
\bibinfo{author}{\bibfnamefont{K.}~\bibnamefont{Fukushima}},  
\bibinfo{author}{\bibfnamefont{M.}~\bibnamefont{Ruggieri}}, 
\bibnamefont{and}
\bibinfo{author}{\bibfnamefont{R.}~\bibnamefont{Gatto}}, 
\bibinfo{journal}{Phys. Rev. D} \textbf{\bibinfo{volume}{81}}, 
\bibinfo{pages}{114031} (\bibinfo{year}{2010}). 
%
\bibitem{Abelev}
\bibinfo{author}{\bibfnamefont{B.}~\bibfnamefont{I.}~\bibnamefont{Abelev et al.~[STAR Collaboration]}},
\bibinfo{journal}{Phys. Rev. Lett.} \textbf{\bibinfo{volume}{103}},
\bibinfo{pages}{251601} (\bibinfo{year}{2009}); 
\bibinfo{journal}{Phys. Rev. C} \textbf{\bibinfo{volume}{81}},
\bibinfo{pages}{054908} (\bibinfo{year}{2010}).  
%
\bibitem{VW}
\bibinfo{author}{\bibfnamefont{C.}~\bibnamefont{Vafa}}
\bibnamefont{and}
\bibinfo{author}{\bibfnamefont{E.}~\bibnamefont{Witten}},
\bibinfo{journal}{Phys. Rev. Lett.} \textbf{\bibinfo{volume}{53}},
\bibinfo{pages}{535} (\bibinfo{year}{1984}). 
%
\bibitem{Dashen}
\bibinfo{author}{\bibfnamefont{R.}~\bibnamefont{Dashen}}, 
\bibinfo{journal}{Phys. Rev. D} \textbf{\bibinfo{volume}{3}},
\bibinfo{pages}{1879} (\bibinfo{year}{1971}). 
%
\bibitem[{\citenamefont{Witten}(1980)}]{Witten}
\bibinfo{author}{\bibfnamefont{E.}~\bibnamefont{Witten}},
\bibinfo{journal}{Ann. \ Phys.} \textbf{\bibinfo{volume}{128}},
\bibinfo{pages}{363} (\bibinfo{year}{1980}). 
%
\bibitem[{\citenamefont{Vecchia and Veneziano}(1980)}]{VV}
\bibinfo{author}{\bibfnamefont{P.}~\bibnamefont{di Vecchia}}, 
\bibnamefont{and}
\bibinfo{author}{\bibfnamefont{G.}~\bibnamefont{Veneziano}},
\bibinfo{journal}{Nucl. \ Phys. } \textbf{\bibinfo{volume}{B171}},
\bibinfo{pages}{253} (\bibinfo{year}{1980}). 
%
\bibitem{Smilga}
\bibinfo{author}{\bibfnamefont{A.}~\bibfnamefont{V.}~\bibnamefont{Smilga}}, 
\bibinfo{journal}{Phys. Rev. D} \textbf{\bibinfo{volume}{59}},
\bibinfo{pages}{114021} (\bibinfo{year}{1999}). 
%
\bibitem{Tytgat}
\bibinfo{author}{\bibfnamefont{M.}~\bibfnamefont{H.}~\bibfnamefont{G.}~\bibnamefont{Tytgat}}, 
\bibinfo{journal}{Phys. Rev. D} \textbf{\bibinfo{volume}{61}},
\bibinfo{pages}{114009} (\bibinfo{year}{2000}). 
%
\bibitem{ALS}
\bibinfo{author}{\bibfnamefont{G.}~\bibfnamefont{Akemann}}, 
\bibinfo{author}{\bibfnamefont{J.}~\bibfnamefont{T.}~\bibnamefont{Lenaghan}}, 
\bibnamefont{and}, 
\bibinfo{author}{\bibfnamefont{K.}~\bibnamefont{Splittorff}}, 
\bibinfo{journal}{Phys. Rev. D} \textbf{\bibinfo{volume}{65}},
\bibinfo{pages}{085015} (\bibinfo{year}{2002}). 
%
\bibitem{Creutz}
\bibinfo{author}{\bibfnamefont{M.}~\bibfnamefont{Creutz}}, 
\bibinfo{journal}{Phys. Rev. Lett.} \textbf{\bibinfo{volume}{92}},
\bibinfo{pages}{201601} (\bibinfo{year}{2004}). 
%
\bibitem[{\citenamefont{Metlitski and Zhitnitsky}(2005)}]{MZ}
\bibinfo{author}{\bibfnamefont{M.}~\bibfnamefont{A.}~\bibnamefont{Metlitski}}, 
\bibnamefont{and}
\bibinfo{author}{\bibfnamefont{A.}~\bibfnamefont{R.}~\bibnamefont{Zhitnitsky}},
\bibinfo{journal}{Nucl. \ Phys. } \textbf{\bibinfo{volume}{B731}},
\bibinfo{pages}{309} (\bibinfo{year}{2005}); 
\bibinfo{journal}{Phys. Lett.\ B} \textbf{\bibinfo{volume}{633}},
\bibinfo{pages}{721} (\bibinfo{year}{2006}). 
%
\bibitem[{\citenamefont{Fujihara et al. }(2010)}]{FIK}
\bibinfo{author}{\bibfnamefont{T.}~\bibnamefont{Fujihara}},
{\bibfnamefont{T.}~\bibnamefont{Inagaki}},
\bibnamefont{and}
\bibinfo{author}{\bibfnamefont{D.}~\bibnamefont{Kimura}},
\bibinfo{journal}{Prog. Theor. Phys.} \textbf{\bibinfo{volume}{117}},
\bibinfo{pages}{139} (\bibinfo{year}{2007}). 
%
\bibitem[{\citenamefont{Boer and Boomsma}(2008{\natexlab{a}})}]{Boer}
\bibinfo{author}{\bibfnamefont{D.}~\bibnamefont{Boer}} 
\bibnamefont{and}
\bibinfo{author}{\bibfnamefont{J.}~\bibfnamefont{K.}~\bibnamefont{Boomsma}},
  \bibinfo{journal}{Phys.\ Rev. D} \textbf{\bibinfo{volume}{78}},
  \bibinfo{pages}{054027} (\bibinfo{year}{2008}). 
%
\bibitem[{\citenamefont{Boomsma and Boer}(2008{\natexlab{a}})}]{Boomsma}
\bibinfo{author}{\bibfnamefont{J.}~\bibfnamefont{K.}~\bibnamefont{Boomsma}}
\bibnamefont{and}
\bibinfo{author}{\bibfnamefont{D.}~\bibnamefont{Boer}}, 
  \bibinfo{journal}{Phys.\ Rev. D} \textbf{\bibinfo{volume}{80}},
  \bibinfo{pages}{034019} (\bibinfo{year}{2009}). 
%
\bibitem[{\citenamefont{Chatteriee et al}(2011)}]{Chatteriee}
\bibinfo{author}{\bibfnamefont{B.}~\bibnamefont{Chatteriee}}, 
\bibinfo{author}{\bibfnamefont{H.}~\bibnamefont{Mishra}} 
\bibnamefont{and}
\bibinfo{author}{\bibfnamefont{A.}~\bibnamefont{Mishra}},  
\bibinfo{howpublished}{arXiv:1111.4061 [hep-ph]}(\bibinfo{year}
{2011}). 
%
\bibitem[{\citenamefont{Kouno et al}(2011)}]{Kouno_CP}
\bibinfo{author}{\bibfnamefont{H.}~\bibnamefont{Kouno}}, 
\bibinfo{author}{\bibfnamefont{Y.}~\bibnamefont{Sakai}},
\bibinfo{author}{\bibfnamefont{T.}~\bibnamefont{Sasaki}}, 
\bibinfo{author}{\bibfnamefont{K.}~\bibnamefont{Kashiwa}}, 
\bibnamefont{and}
\bibinfo{author}{\bibfnamefont{M.}~\bibnamefont{Yahiro}},  
\bibinfo{journal}{Phys.\ Rev.\  D} \textbf{\bibinfo{volume}{83}},
\bibinfo{pages}{ 076009} (\bibinfo{year}{2011}). 
%
\bibitem[{\citenamefont{Sakai et al}(2011)}]{Sakai_CP}
\bibinfo{author}{\bibfnamefont{Y.}~\bibnamefont{Sakai}},
\bibinfo{author}{\bibfnamefont{H.}~\bibnamefont{Kouno}}, 
\bibinfo{author}{\bibfnamefont{T.}~\bibnamefont{Sasaki}}, 
\bibnamefont{and}
\bibinfo{author}{\bibfnamefont{M.}~\bibnamefont{Yahiro}},  
\bibinfo{journal}{Phys.\ Lett.\  B} \textbf{\bibinfo{volume}{705}},
\bibinfo{pages}{349} (\bibinfo{year}{2011}). 
%
\bibitem[{\citenamefont{Sasaki et al.}(2012)}]{Sasaki}
\bibinfo{author}{\bibfnamefont{T.}~\bibnamefont{Sasaki}}, 
\bibinfo{author}{\bibfnamefont{J.}~\bibnamefont{Takahashi}}, 
\bibinfo{author}{\bibfnamefont{Y.}~\bibnamefont{Sakai}},
\bibinfo{author}{\bibfnamefont{H.}~\bibnamefont{Kouno}},
\bibnamefont{and} 
\bibinfo{author}{\bibfnamefont{M.}~\bibnamefont{Yahiro}}, 
\bibinfo{journal}{Phys. Rev.\ D} \textbf{\bibinfo{volume}{85}},
\bibinfo{pages}{056009} (\bibinfo{year}{2012}); 
%
\bibitem[{\citenamefont{Kobayashi and Maskawa}(1970)}]{KMK}
\bibinfo{author}{\bibfnamefont{M.}~\bibnamefont{Kobayashi}}, 
\bibnamefont{and}
\bibinfo{author}{\bibfnamefont{T.}~\bibnamefont{Maskawa}},
\bibinfo{journal}{Prog. Theor. Phys. } \textbf{\bibinfo{volume}{44}},
\bibinfo{pages}{1422} (\bibinfo{year}{1970});
\bibinfo{author}{\bibfnamefont{M.}~\bibnamefont{Kobayashi}}, 
\bibinfo{author}{\bibfnamefont{H.}~\bibnamefont{Kondo}},
\bibnamefont{and}
\bibinfo{author}{\bibfnamefont{T.}~\bibnamefont{Maskawa}},
\bibinfo{journal}{Prog. Theor. Phys. } \textbf{\bibinfo{volume}{45}},
\bibinfo{pages}{1955} (\bibinfo{year}{1971}). 
%
\bibitem[{\citenamefont{Frank}(2003)}]{Frank}
\bibinfo{author}{\bibfnamefont{M.}~\bibnamefont{Frank}}, 
\bibinfo{author}{\bibfnamefont{M.}~\bibnamefont{Buballa}}, 
\bibnamefont{and} 
\bibinfo{author}{\bibfnamefont{M.}~\bibnamefont{Oertel}}, 
\bibinfo{journal}{Phys. Lett.\ B} \textbf{\bibinfo{volume}{562}},
\bibinfo{pages}{221} (\bibinfo{year}{2003}).
%
\bibitem[{\citenamefont{Hansen}(2007)}]{Hansen}
\bibinfo{author}{\bibfnamefont{H.}~\bibnamefont{Hansen}}, 
\bibinfo{author}{\bibfnamefont{W.}~\bibfnamefont{M.}~\bibnamefont{Alberio}}, 
\bibinfo{author}{\bibfnamefont{A.}~\bibnamefont{Beraudo}}, 
\bibinfo{author}{\bibfnamefont{A.}~\bibnamefont{Molinari}}, 
\bibinfo{author}{\bibfnamefont{M.}~\bibnamefont{Nardi}}, 
\bibnamefont{and} 
\bibinfo{author}{\bibfnamefont{C.}~\bibnamefont{Ratti}}, 
\bibinfo{journal}{Phys. Rev.\ D} \textbf{\bibinfo{volume}{75}},
\bibinfo{pages}{065004} (\bibinfo{year}{2007}).
%
\bibitem[{\citenamefont{Andersen}(2010)}]{Andersen}
\bibinfo{author}{\bibfnamefont{J.}~\bibfnamefont{O.}~\bibnamefont{Andersen}}, 
\bibinfo{author}{\bibfnamefont{L.}~\bibfnamefont{T.}~\bibnamefont{Kyllingstad}}, 
\bibnamefont{and} 
\bibinfo{author}{\bibfnamefont{K.}~\bibnamefont{Splittroff}}, 
\bibinfo{journal}{J. High Energy Phys. } \textbf{\bibinfo{volume}{01}},
\bibinfo{pages}{055} (\bibinfo{year}{2010}); 
%
\bibitem[{\citenamefont{Sasaki et al.}(2010)}]{Sakai_sign}
\bibinfo{author}{\bibfnamefont{Y.}~\bibnamefont{Sakai}},
\bibinfo{author}{\bibfnamefont{T.}~\bibnamefont{Sasaki}}, 
\bibinfo{author}{\bibfnamefont{H.}~\bibnamefont{Kouno}},
\bibnamefont{and} 
\bibinfo{author}{\bibfnamefont{M.}~\bibnamefont{Yahiro}}, 
\bibinfo{journal}{Phys. Rev.\ D} \textbf{\bibinfo{volume}{82}},
\bibinfo{pages}{096007} (\bibinfo{year}{2010}); 
%

\end{thebibliography}
\end{document}